\documentstyle[11pt,newpasp,twoside]{article}
\begin{document}
\title{Protoneutron stars and neutron stars} \author{Gondek-Rosi\'nska D.,
  Haensel P., Zdunik J. L.}  \affil{Nicolaus Copernicus Astronomical Center,
  Polish Academy of Sciences, Bartycka 18, 00-716 Warszawa, Poland}

\begin{abstract}
We find constraints on minimum and maximum mass of ordinary neutron
stars imposed by {}
their early evolution (protoneutron 
star stage).  We calculate  {}  
models of protoneutron stars  using  a 
realistic standard equation of state of hot, dense matter valid for both 
supranuclear and subnuclear densities. Results for different values of 
the nuclear incompressibility are presented.


\end{abstract}

\section{Introduction}

The newborn neutron star (protoneutron star - PNS) does not resemble the cold 
ordinary neutron star (NS). It is a very hot and lepton rich object. The 
minimum and the maximum allowable mass of PNSs are {} different than that {} 
corresponding to cold NSs. PNS evolves either to BH or to stable NS 
depending on its total number of baryons. 

  
Constraints on {\bf $M_{\rm max}$} of NSs imposed by composition and
equation of state (EOS) of hot dense stellar interior were studied by
numerous authors (e.g.  Takatsuka 1995, Bombaci et. al 1995, Bombaci
1996, Prakash et al.  1997). For low densities they took EOS of
cold matter. Here we use a unified dense
matter model, which holds for both supranuclear and subnuclear
densities. We constrain {\bf $M_{\rm max}$} and 
{\bf $M_{\rm min}$} of NSs assuming conservation of the total baryon number 
of the star during its evolution.  Since accretion on the forming protoneutron star ceases few seconds after core bounce
 (Chevalier 1989) we restrict ourselves to
the late stage PNS (see Goussard et al. 1998, Strobel et al. 1999
for the discussion and definitions of the stages of PNS evolution);
 note that Strobel at al. (1999) considered PNS 
about 50-100 ms after core bounce as a initial stage. 
The problem of $M_{\rm min}$ of
NS was studied by Gondek et al. 1998, but only for  moderately 
stiff equation of state. Here we present results also for stiffer and softer
equation of state.

\section{Calculations and results}
A few seconds after
birth $t \simeq 2-4 \ {\rm s}$ the matter in the core of a hot neutron star has
almost constant lepton fraction ($Y_l=0.3-0.4$) and entropy per baryon
($s=1-2$ in unit of the Boltzmann constant $k_B$) (Burrows \& Lattimer
1986). Our models of PNS are composed of a hot, neutrino-opaque interior, 
and neutrino-transparent envelope. 
We find the position of the neutrinosphere in a
self-consistent way (for more details see Gondek et al. 1997). 
We consider the hot isentropic interior with $s=1$ or
$s=2$ and a significant trapped lepton number $Y_l=0.4$. 
All calculations were done for a realistic, standard Lattimer-Swesty (1991) hot, 
dense matter model. To cover the present uncertainties of nuclear interaction 
and many-body technique, we construct models of PNS for three different values
of the nuclear incompressibility; $K =$ 375, 220, 180 MeV.
We consider PNS as an isolated, non-rotating, spherically
symmetric object, which has no magnetic field. Under these assumptions the
structure of the star is solved numerically using TOV equations
(Tolman 1939, Oppenheimer \& Volkoff 1939) for a given EOS. 

The gravitational masses of
NSs determined by maximum and minimum allowable masses of PNSs are smaller than 
$\sim 2.5 M_\odot$ and greater than $\sim 0.5M_\odot$ (this is by a factor 
of ten larger than the minimum masses of static NSs sequences). 
The minimum baryon mass of PNSs at the earlier stages of their evolution
was found to be 2-3 times larger (Goussard et al. 1998, 
Strobel et al. 1999 for the EOS of dense hot matter based on 
the Thomas-Fermi model) than in the stage considered by us. 
In this case a hot shocked stellar envelope has main influence on the 
minimum mass of PNS sequence.

We show that the maximum critical gravitational mass of a NS is by 
$\sim 0.14-0.19 M_\odot$ (depending on the nuclear incompressibility $K$)
smaller than maximum mass of static NSs sequence. This result seems to be
characteristic for a standard model of dense matter, composed of
nucleons and leptons (e.g. Takatsuka 1995, Bombaci 1996).
The conclusions are different when we take into account existence of
an exotic components (hyperons, pions or kaons condensate or quark
matter) at hight densities (Bombaci et al. 1996). Appearance of
exotic matter could dramatically soften the EOS of dense matter 
lowering maximum allowable mass of NSs (e.g. Reddy et al. 1999). 
The deleptonization and cooling of PNSs of baryon mass
exceeding the maximum allowable baryon mass for NSs, would lead to
their collapse into BH.

\acknowledgments 
This work was supported in part by the following
grants KBN-2P03D01413 and KBN-2P03D02117.


\begin{references}
\reference Baumgarte T.W., Shapiro S.L., Teukolsky S.A., 1996, ApJ 458, 680
\reference Bombaci I., Prakash M., Prakash M., Ellis P.J.,
Lattimer J.M., Brown G.E., 1995, Nucl. Phys. A, 583, 623 
\reference Bombaci I., 1996, A\&A 305, 871 
\reference Brown G.E., Bethe H.A., 1994, ApJ 423, 659 
\reference Burrows A., Lattimer J.M., 1986, ApJ, 307, 178 
\reference Chevalier R., 1989, ApJ, 346, 847
\reference Gondek D., Haensel P., Zdunik J. L., 1997, A\&A, 325, 217 
\reference Gondek D., Haensel P., Zdunik J. L., 1998, \pasp, 138, 131, astro-ph/0012541
\reference Goussard J-O., Haensel P., Zdunik J. L., 1998, A\&A, 330, 1005 
\reference Lattimer J.M., Swesty F.D., 1991, Nucl. Phys. A535, 331 
\reference  K. Strobel, C. Schaab, M. Weigel, 1999, A\&A, 350, 497
\reference Takatsuka T., 1995, Nucl. Phys. A588, 365c 
\end{references}
\end{document}